\documentstyle[preprint,aps]{revtex}

\global\arraycolsep=2pt
\makeatletter
\def\fmslash{\@ifnextchar[{\fmsl@sh}{\fmsl@sh[0mu]}}
\def\fmsl@sh[#1]#2{  \mathchoice
    {\@fmsl@sh\displaystyle{#1}{#2}}
           {\@fmsl@sh\textstyle{#1}{#2}}
                 {\@fmsl@sh\scriptstyle{#1}{#2}}
                       {\@fmsl@sh\scriptscriptstyle{#1}{#2}}}
\def\@fmsl@sh#1#2#3{\m@th\ooalign{$\hfil#1\mkern#2/\hfil$\crcr$#1#3$}}
\makeatother

\begin{document}
\draft

\title{Relaxation and screening in Auger emission: 
an explanation for the changes from bandlike to quasi-atomiclike 
CVV Auger spectra across the transition metal series}
\author{Jianmin Yuan}
\address{Department of Applied Physics, 
National University of Defense Technology, 
Changsha 410073, P. R. China}
\date{\today}
\maketitle

\thispagestyle{empty}

\begin{abstract}
Supercell method is used to study the relaxation and screening effects 
during the Auger transition in metals. Our consideration is based 
on the assumption that when a core-hole 
exists long enough before the Auger transition occurs, the occupied valence
states relax to screen the core-hole which results in a redistribution of
the valence electrons, in particular within the atom that contains the
core-hole. In order to make the interaction between the core-holes sited 
at different atoms negligible, the real metal 
is simulated by supercells repeated 
periodically. In each supercell one atom is considered to 
have a core-hole and many others without core-hole. 
The electronic states concerned by the Auger transition are
calculated by the self-consistent full-potential linearized augmented plane
wave (FLAPW) method. 
The occurrence of both bandlike and quasi-atomiclike Auger spectra across 
transition metal series is related clearly to the different 
responses of these metals to the existence of a core-hole 
depending on whether their d-bands are partially or completely filled. As 
examples, L$_3$VV and M$_1$VV Auger spectral profiles of Cu have been 
calculated in a reasonably well agreement with experiment.
\end{abstract}


\pagenumbering{roma}

\vspace{4mm}

\newpage \pagenumbering{arabic}

\section{Introduction}
The idea of correlating Auger spectral 
line shapes from elemental metals with their
electronic structure goes back to Lander\cite{lander}. It was shown that 
the Auger spectral line shape of a core-valence-valence (CVV) transition 
is determined by the self-convolution of the local density of 
(valence) states (LDOS) if the state dependence of the 
transition matrix elements may to a good  approximation be ignored 
across the valence band. From the late 1970's, efforts have been given 
to consider the actual variation of the matrix elements and their 
dependence on the angular momenta of the valence 
electrons\cite{feibelman,jennison,davies,weinberger,almbladh,liegener}. 
The results showed that, 
based on the basic assumption of Lander with adequate consideration of the 
state variation of the matrix elements, 
reasonably good agreement between theories and experiments can be obtained 
for the CVV Auger spectral line form of alkali and alkaline-earth metals 
and silicon. Even the spin polarization of the Auger electron can be 
described adequately by the theory\cite{yuan}. 
The basic feature of the CVV Auger spectra of these materials 
is a relatively broad peak reflecting the fact that there is a definite 
width for the valence band.

In contrast to the relatively prosaic CVV Auger spectral line shape of 
alkali and alkaline-earth metals, it is obvious from the experimental 
findings that the Auger spectra of some
materials display quasi-atomic features on a background reflecting the
presence of itinerant states\cite{powell}. The occurrence of quasi-atomiclike 
CVV Auger spectra is typical of some d-band materials\cite{weightman}. 
Relying on this kind of experimental findings, 
Cini\cite{cini} and Sawatzky\cite{sawatzky} developed a theory of
Auger emission based on a Hubbard model. The original form of their theory 
is only applicable to filled band system and has been extended by 
many author to treat the unfilled band and multi-band 
cases\cite{treglia,kotrla,cini92,nolting,verdozzi}. 
In his theory Sawatzky\cite{sawatzky} is
led to the conclusion that when the effective Coulomb interaction $U$ in the
Hubbard Hamiltonian is very large compared to the width of the valence band $%
W$, that is if $U\gg 2W$, the Auger spectra will display a relatively sharp
quasi-atomiclike structure. The parameter $U$ was determined empirically from the
experiments\cite{antonides,yin,victora,kaurila}. 
The obtained values increase from light
(V) to heavy (Zn) 3d-transition metals. The relatively large $U$ for Cu and
Zn is consistent with the fact that these metals show quai-atomiclike sharply
split structures in their valence Auger spectra, while Cr, Fe, and Ni
display relatively broad bandlike features. 
The most interesting
thing is that the change of the CVV Auger spectral line shape from
bandlike to quasi-atomiclike characteristics across the
transition metal series from left to right in the periodic table
occurs dramatically at Cu and Ag, respectively,
for 3d and 4d transition metals, rather than gradually 
from left to right across the whole series. 
Nevertheless, the theory of Cini\cite{cini} and Sawatzky\cite{sawatzky} 
cannot give a quantitative prediction that the 
quasi-atomiclike CVV Auger spectra just start from Cu for the 3d transition 
metals and from Ag for the 4d transition metal, from which the d-band starts 
to be filled completely. 

We will show by an {\sl ab initio} calculation why the quasi-atomiclike 
CVV Auger spectra start from Cu and Ag respectively for the 
3d and 4d series. The relaxation and screening effects are the main factors 
in our consideration. The relaxation and screening effects has been 
investigated by many authors since the late 1970's. In the 
early study of Barth and 
Grossmann\cite{barth}, considerable changes in the LDOS caused by the 
relaxation and screening effects was predicted by means of an empirical 
model. However, after taking the dynamical screening into account, the 
static screening around the core-ionized site has less influences on 
the spectra profiles of the x-ray emission and CVV Auger transition of the 
simple metals\cite{barth,barth2,almbladh}, resulting in the so-called 
final-state rule. 
The effects of screening on the Auger spectral line shape of metallic 
Mg was studied by Davies {\sl et al.}\cite{davies} also using an 
empirical method. Cluster method and transition state were applied by 
Kucherenko\cite{kucherenko} to the investigation of the influence of 
relaxation effects on the shape of the Auger spectra of metals.  
A modified excited atoms model was used to predict 
relaxation and Auger parameter shifts between free atoms and elemental 
solids for Na and Mg\cite{cole}. 
The screening effect was also considered 
by Yuan {\sl et al.}\cite{yuan} to find it's influence on the 
spin polarization of Auger electrons from K and Cr metals. Most recently, 
Weightman\cite{weightman3} discussed the screening and correlation 
effects in CVV and CCV Auger processes of Si. The localized nature of the 
final states and the delocalized nature of the screening in 
semiconductors have influences on both CVV and CCV Auger spectra profiles. 
In the present 
work, relaxation and screening effects will be studied by using a  
supercell method and a definite relation will 
be found between the occurrence of the bandlike or quasi-atomiclike 
CVV Auger spectra and the qualitatively different response of the 
valence band electrons to screen the core-hole. As examples, 
L$_3$VV and M$_1$VV Auger spectral profiles of Cu will be calculated 
to see to what extent the present theory can reproduce experimental 
observations.

\section{Theoretical method}
Both the initial and final Auger states are excited (mutually degenerate) 
N-electron states that can be described within a generalized density 
functional theory\cite{fritsche}. The key idea of density functional 
theory resides in mapping the interacting N-electron system onto that of 
a noninteracting N-electron system having the same spin-resolved  
one-particle densities as the original one but moving in a modified 
external potential. The noninteracting wave function has the form of a 
Slater determinant. The approach used in the following is based on the 
assumption that the transition matrix elements describing the Auger 
process can, to a good approximation, be calculated by using the 
pertinent initial- and final-state determinants instead of the true 
N-electron wave function. As a consequence, the following calculations 
concern only one-electron states by which these Slater determinants 
differ. As for the electronic states of 
the metals we are considering, we employ the
full-potential linearized augmented plane wave (FLAPW) code WIEN95\cite{wien95} 
to calculate the itinerant valence and semi-core
solutions to the Kohn-Sham equations. The FLAPW method requires a
subdivision of the crystal into sufficiently large, but non-overlapping
concentric spheres (atomic spheres) around the atomic nuclei and an
interstitial region between these spheres. Inside the atomic sphere the
one-electron state of band index $n$ is given by 
\begin{equation}
\psi_{n}(\epsilon_{n},\vec{k},\vec{r}) =
\sum\limits_{L}\sum\limits_{\nu=0}^{1} c_{L\nu}^{(n)}(\epsilon_{n},\vec{k}%
)R_{l\nu}(\epsilon_{l},r)Y_{L}(\hat{\vec{r}}) \chi_{\sigma_{n}}
\end{equation}
where spherical harmonics are denoted by $Y_{L}(\hat{\vec{r}})$, $L=(l,m)$.
The quantity $\chi_{\sigma_{n}}$ represents a unit spinor for the spin
orientation $\sigma_{n}=\pm1$. The function $R_{l0}(\epsilon_{l},r)$ is
regular at the origin and solves the radial part of the Kohn-Sham-type
equation for $E=\epsilon_{l}$, $R_{l1}(\epsilon_{l},r)$ denotes its
normalized energy derivative.

The Auger transition rate $P_{fi}^{\sigma_{d}}$ can be cast into the Golden
Rule form\cite{chattarji} 
\begin{equation}
P_{fi}^{\sigma_{d}}(a,d)\propto \sum\limits_{\stackrel{\vec{k}^{\prime},\vec{%
k},n^{\prime},n}{\sigma_{a},m_{s_{a}}}} \mid
M_{fi}^{(\sigma_{d},\sigma_{a},m_{s_{a}})} (\vec{k}^{\prime}n^{\prime};\vec{k%
},n) \mid^{2} \delta (\epsilon_{d}-\epsilon_{n^{\prime}}(\vec{k}%
^{\prime})-\epsilon_{n}(\vec{k})+\epsilon_{a}+\Delta\epsilon (
\vec{k}^{\prime}n^{\prime};\vec{k},n))
\end{equation}
where $\sigma_{d}=\pm1$ refers to the two spin orientations of the outgoing
electron. The spin quantum number of the core-hole spinor state $\psi_{a}(%
\vec{r})$ is denoted by $m_{s_{a}}$ ($=\pm\frac{1}{2}$) referring to the
total angular momentum $j_{a}=l_{a}+m_{s_{a}}$ of the spin-orbit split
core-hole doublet states. $\Delta\epsilon (
\vec{k}^{\prime}n^{\prime};\vec{k},n)$ represents the non-averaged 
interaction between the two holes after the Auger 
transition occurs\cite{sawatzky}. It can 
be calculated by using the wave functions of the two concerned states:
\begin{equation}
\Delta\epsilon (\vec{k}^{\prime}n^{\prime};\vec{k},n)=
\langle \psi_{b}^{\ast}(\vec{r}_{1})\psi_{c}^{\ast}(\vec{r}_{2})
\left|\frac{1}{|\vec{r}_{1}-\vec{r}_{2}|}\right|
\psi_{b}(\vec{r}_{1})\psi_{c}(\vec{r}_{2})\rangle+
\delta_{\sigma_{b},\sigma_{c}}EX
\end{equation}
where $\psi_{b}(\vec{r})$ is given by Eq.(1). $EX$ represents the exchange 
part. By using the expansion given in Eq.(1), the above element can be 
expressed in details as:
$$
\Delta\epsilon (\vec{k}^{\prime}n^{\prime};\vec{k},n)=
\sum\limits_{L_{b}}\sum\limits_{L_{c}}\sum\limits_{L_{b}^{\prime}}
\sum\limits_{L_{c}^{\prime}}\sum\limits_{L}\frac{4\pi}{2l+1}
G(L_{b}^{\prime},L_{b},L)G(L_{c},L_{c}^{\prime},L) 
$$
\begin{equation}
\times\sum\limits_{\nu_{b}=0}^{1}\sum\limits_{\nu_{c}=0}^{1}
\sum\limits_{\nu_{b}^{\prime}=0}^{1}
\sum\limits_{\nu_{c}^{\prime}=0}^{1}
c_{L_{b}\nu_{b}}^{b \ast}c_{L_{c}\nu_{c}}^{c \ast}
c_{L_{b}^{\prime}\nu_{b}^{\prime}}^{b}
c_{L_{c}^{\prime}\nu_{c}^{\prime}}^{c}
F_{\nu_{b}\nu_{c}\nu_{b}^{\prime}\nu_{c}^{\prime}}
(l_{b},l_{c},l_{b}^{\prime},l_{c}^{\prime})+
\delta_{\sigma_{b},\sigma_{c}}EX
\end{equation} 
in which $G(L_{b}^{\prime},L_{b},L)$ represent the Gaunt integrals 
and $F_{\nu_{b}\nu_{c}\nu_{b}^{\prime}\nu_{c}^{\prime}}$ is shorthand 
for
$$
F_{\nu_{b}\nu_{c}\nu_{b}^{\prime}\nu_{c}^{\prime}}
(l_{b},l_{c},l_{b}^{\prime},l_{c}^{\prime})=
\int_{0}^{r_{0}}R_{l_{b}\nu_{b}}(r_{1})
R_{l_{b}^{\prime}\nu_{b}^{\prime}}(r_{1})
\left [\frac{1}{r_{1}^{l+1}}\int_{0}^{r_{1}}
r_{2}^{l+2}
R_{l_{c}\nu_{c}}(r_{2})R_{l_{c}^{\prime}\nu_{c}^{\prime}}(r_{2})dr_{2}
\right.
$$
\begin{equation}
\left.
+r_{1}^{l}\int_{r_{1}}^{r_{0}}\frac{1}{r_{2}^{l-2}}
R_{l_{c}\nu_{c}}(r_{2})R_{l_{c}^{\prime}\nu_{c}^{\prime}}(r_{2})dr_{2}
\right ]dr_{1}.
\end{equation}
With CVV Auger transitions one is dealing with a core-hole state that is, 
to a good approximation, confined to the pertaining atom. The intra-atomic 
transition, in which the final two holes located in the same atom with 
the initial core-hole, contributes the dominant part to the Auger spectra. 
Hence the integration over the $r$-dependent functions may be restricted 
to the atomic sphere of that atom without causing considerable errors. 
The matrix elements on the right-hand side of
Eq.(2) can be split into two portions 
\begin{equation}
M_{fi}^{(\sigma_{d},\sigma_{a},m_{s_{a}})}(\vec{k}^{\prime},n^{\prime};\vec{k%
},n)=
D^{(m_{s_{a}})}_{abcd}\delta_{\sigma_{a},\sigma_{b}}\delta_{\sigma_{c},%
\sigma_{d}}
-E^{(m_{s_{a}})}_{abcd}\delta_{\sigma_{a},\sigma_{c}}\delta_{\sigma_{b},%
\sigma_{d}}
\end{equation}
where $D^{(m_{s_{a}})}_{abcd}$ and $E^{(m_{s_{a}})}_{abcd}$ denote,
respectively, the so-called direct and exchange portion of the transition
matrix element. The details for the calculations of the transition 
matrix element can be found in our previous paper\cite{yuan}. 

\section{Result and discussion}

It is well known that the CVV Auger spectra of transition metals with
partially filled d-bands like Cr, 
Fe, Ni and Pd etc. show relatively broad
bandlike features, while the spectra of transition metals with completely
filled d-bands like Cu, Zn and Ag are 
conspicuous by sharp quasi-atomiclike
structures. The approach used by Cini\cite{cini} 
and Sawatzky\cite{sawatzky}
to explain the quasi-atomiclike Auger spectra is based on a Hubbard-type
Hamiltonian that contains the parameter $U$ describing the effective Coulomb
interaction and another parameter $W$ referring to the band width. If $%
U\gg 2W$, it was shown by these authors that the two-hole density of states
splits into a broad and a narrow structure. The occurrence of the latter is
correlated with the observed quasi-atomiclike Auger peaks. By exploiting certain
experimental findings it can be made evident that Cu and Zn are associated
with a larger ratio $U/2W$ compared to the other 3d-transition
metals.

The theory of Cini\cite{cini} and Sawatzky\cite{sawatzky} 
can qualitatively explain the change from bandlike to
quasi-atomiclike Auger emission as one goes from the light to the heavy
3d-transition metals. Nevertheless, due to the model Hamiltonian and the 
empirical parameters involved in their theory, they cannot predict 
in a sense of an {\sl ab initio} way that 
the change of the CVV Auger spectrum from bandlike to quasi-atomiclike 
features occurs just from Ni to Cu\cite{fuggle} rather than somehow earlier or 
later in the 3d transition series by noting the fact that the ratio 
$U/2W$ increases monotonously from Fe to Zn\cite{antonides}. 
We have therefore reexamined this
problem from a first-principles point of view. Our considerations are based
on the two-step model of the Auger electron emission. An initial core-hole
is first generated by the absorption of an incoming photon or by impact with
some other particle. If that core-hole exists long enough before the Auger
transition occurs, the occupied valence states relax to screen the core-hole
which results in a redistribution of the valence electrons, in particular
within the atom that contains the core-hole.

In our theoretical treatment, a core-hole state can be obtained by simply 
putting one less electron in the considered core orbital. In order to make 
the interactions between the core-hole negligible, 
supercell method is employed, in which an enough large unit cell 
containing one atom with a core-hole and many other atoms without 
core-hole is repeated periodically, and we take electronic structure 
calculations with many chemically same atoms per unit cell. In our 
calculations, except for the unit cell being chosen much larger than 
the usual one there is not any change made on the real crystal 
structure, and except for one less electron being putted in a  
core orbital of a specific atom, there are not any other extra 
environmental conditions taken on the supercell. Therefore, after an 
enough large unit cell is used, 
we carry out an {\sl ab initio} calculation. 
The accuracy of the calculation is examined by having the supercell 
results being the same as the simple crystal calculation when the 
core-hole state does not exist. We take the atom having the core-hole 
at the center of the unit cell, and take the unit cell large enough 
to have the LDOS of the atoms at the corner of the cell being 
nearly the same of the simple crystal. 
By this way, we have, in a unit cell, up to 
sixteen atoms for the Bcc and Hcp metals and thirty two atoms for 
Fcc crystals. The unit cell is a simply cubic one for Bcc and Fcc crystals, 
and a hexagonal one for the Hcp cases. 

We first confine ourselves to describe the relaxed electron
distribution after the core-hole excitation. The characteristics of the 
valence electronic states can be seen from the LDOS of the 
concerned atoms. The bandlike or quasi-atomiclike 
features of the Auger spectra can be extracted from the LDOS  
of the structural calculations. 
The valence band was appropriately filled
higher up to ensure charge neutrality of the system. This is just 
equal to exciting the core electron to the valence band instead of the 
vacuum states. From Figs. 1 to 9, we show the LDOS 
obtained for Sc, Ti, Cr, Fe, Ni, Cu, Zn, 
Pd and Ag to see the changes of the LDOS induced by the existence of a 
core-hole from light to heavy elements across the 3d transition 
series, in particular the changes of the LDOS from Ni to Cu and 
from Pd to Ag as examples of the transition between the two kind 
responses to screen the core-hole. 
The plots refer to the density of states of the pertinent atom with a
core-hole by the solid lines, to that of the atoms at the corner of the 
unit cell without the core-hole by the dot lines, and to the LDOS of 
the perfect crystals by the long-short dashed lines. 
Obviously, from Sc to Ni and Pd the relaxation only changes the 
distribution of the density of states within the valence band, 
although much higher values of the LDOS are 
observed at the bottom of the d-bands for Ni and Pd, which will result 
in a relatively narrow Auger line shape. In contrast to the above metals, 
for Cu, Zn and Ag the relaxation leads
to a sharply peaked structure strongly shifted away from the original d-band
complex. While the latter contains sizable p-type portions, the peaked
structure is almost purely d-type. The position of the sharp structure falls
into the gap between the onset of the s-p valence band and the lower lying
3p-core level band. If one could gradually enlarge the unit cell, the peak
would more and more attain the form of a delta-function reflecting the fact
that one is dealing with a d-type quasi-atomiclike impurity state. Therefore, the
Auger transition closely resembles that of a free atom. The quasi-atomiclike
splitting one observes with the respective peak can hence easily be
understood as a final state effect that results from the coupling in the
open 3d-shell after the Auger transition has taken place. Spin polarized 
calculations are carried out for Cr, Fe and Ni to consider the 
antiferromagnetic and ferromagnetic properties of these metals.

The difference response of the valence electrons to the relaxation effects 
comes from the different occupations of the screening electrons to s and 
p or d orbitals. For the full filled d-band metals, the screening 
electrons can only go to the s  and p orbitals, which lies above the 
d-band, and the strong core potential on the d-band cannot be screened 
effectively. Due to the strong attractive core potential, the d-band 
moves towards a much lower energy and display a typical localized state 
characteristics. For the partially filled d-band metals, the screening 
electrons mainly occupy the empty d orbitals resulting in a more effective 
screening on the core attractive potential for the d-band, and the 
band structure of the d electrons remains unchanged with a definite 
width of the band. As examples, we consider the changes of the valence 
charge inside the atomic sphere caused by relaxation for Pd and Ag. 
The valence charge inside the atomic sphere with the core-hole increases 
1.003 and 0.718 for Pd and Ag respectively. The increased charge of 
1.003 of Pd distributes 0.039 in s-wave, 0.042 in p-wave, and 
0.921 in d-wave, while the increased charge of 0.718 of Ag has 
0.150 in s-wave, 0.159 in p-wave and 0.408 in d-wave. The results 
support the fact that for Pd most of the increased charge to screen 
the core attractive potential comes from the d partial waves, while 
for Ag most of the increased charge to screen 
the core attractive potential comes from the s and p partial waves 
and the main part is outside the atomic sphere. Because of the very 
small DOS at the Fermi energy, the screening radius\cite{ziman} of Ag is much 
larger than that of Pd, i.e. we have the delocalized nature of 
screening for the full filled d-band transition metals. The increased d partial 
charge inside the atomic sphere for Ag is mainly attributed to the 
contraction of the d orbital caused by the strong core attractive 
potential.   

As examples, we also calculated the L$_3$MM and M$_1$VV 
Auger spectra of Cu as 
described by the Golden Rule of Eq.(2). In the calculation, if we neglect 
the Coulomb interaction between the two final holes, we will have an 
$\delta$-function type spectra as a result of the self-convolution of 
the LDOS shown in Fig. 6 by the solid line. 
In Figs. 10 and 11, we give the calculated Auger 
spectra by including the $\Delta\epsilon$ in Eq.(2). The result shows 
some structures that is similar to those obtained by the 
experiments\cite{sarma,madden,weightman2}. The  
structures and their separations observed and labeled as B, D and E by 
Madden {\sl et al.}\cite{madden} 
for the L$_3$VV transition are reproduced quite 
well by our calculations. Difference in the details of the relative height 
of the peaks B and E can be found between our theory and the 
experiment of Madden {\sl et al.}\cite{madden}, but the relative 
intensities of B and E agree much better with the latest measured result of
Sarma {\sl et al.}\cite{sarma} with subdued satellites  
in the left side of the main peak D.  
Please be noted that the L$_3$VV and M$_1$VV 
spectra are obtained with different electronic wavefunctions, which 
are calculated by using the supercell model with a core-hole, respectively, in 
L$_3$ and M$_1$ core orbitals. Apparent difference between the 
L$_3$VV and M$_1$VV spectral profiles are revealed by both the present 
theory and the experiment\cite{madden}. 

The Auger spectra obtained without the screening 
effects is also presented in Figs. 10 and 11 by the dashed lines. 
It is obvious
that the much more broader line shape cannot reproduce the experimental 
observations. The main features of the L$_3$MM and M$_1$VV 
Auger spectra of Cu is the relatively sharp 
peaks, which is called atomiclike and corresponds to 
the localized characteristics of the concerned electron states. 
The separated peaks, for example, of the L$_3$MM Auger spectra is the 
multiplet split caused by the hole-hole interaction in the final state. 
With a core-hole, 
the localization of the initial states for the full filled d-band transition 
metals have been seen clearly from the LDOS in the 
Figs. 6,7 and 9. The position of the localized initial states are well below the 
normal valence band.  After the Auger transition, 
there is not hole in the core, but there are 
two holes in the final states. If the dynamical screening is as  
significant as in the simple sp metals, the final two hole states should 
be very close to the valence electron states of the ground metal. 
Nevertheless, the dynamical screening in the full filled d-band 
transition metals is much less effective than that in the simple metals.  
The reason will be given in the following discussions. Therefore, the 
final two hole states of the CVV Auger transition of Cu should be 
much more localized compared with the ground valence states. 
In order to simplify the calculations( with orthogonal orbitals), 
we use the initial states, in a good approximation, to replace the true 
final orbitals. The multiplet split caused by the hole-hole 
interaction in the final state is also taken into account in the present 
treatment by including the $\Delta\epsilon$ term in Eq.(2), and  
the present calculations are not exactly 
one-electron model in the normal sense. 
Even in the normal one-electron model, one usually does not give the 
same results with and without the core-hole screening. According to 
Lander\cite{lander}, the Auger spectral line shape of a CVV Auger 
transition is determined by the self-convolution of the LDOS.  
We have shown that the LDOS's of the full filled d-band transition 
metals with and without the core-hole screening are dramatically 
different. The more broader line shape of the dashed lines in 
Figs. 10 corresponds to the much wider valence band of  
the perfect crystal, and the relatively higher intensity of the 
right peak is also consistent with the shape of the LDOS. 
We do not give the absolute values of the kinetic 
energy of the Auger electron, because the local-density-approximation (LDA) 
cannot predict reliable energy distance between core state and valence 
band. 

In the present study, we did not consider the competing of the Auger 
emission to the screening because of the fact that in most metals the 
valence screening is a faster process than a CVV Auger emission. We 
did not consider the dynamical screening either. The dynamical 
screening would most likely have some influences on the Auger 
emission spectra. In some previous studies of the dynamical screening 
effect\cite{barth,barth2,almbladh}, final state rule had been arrived 
for simple metals by using a dynamical approach\cite{nozieres}. 
The rule states that the CVV Auger spectral profiles of the simple 
metals are most determined by the LDOS of the ground state valence 
band and that the screening effect around the core-hole state before 
the Auger transition are compensated significantly by the dynamical 
screening. Nevertheless, this conclusion derived for simple metals 
cannot be applied to the transition metals without modifications, 
because some of the transition metals, for examples Cu and Zn, have 
quite different valence band structure. For these transition metals, 
the density of state at the Fermi level has very small values compared 
with the simple metals and other transition metals with partially filled 
d-band. Therefore, the screening behavior of these transition metals 
is somehow like to that in semiconductors, which results in  
localized final states and delocalized 
screening\cite{weightman3,ziman} on the core-hole 
before the Auger transition and on the valence double hole after 
the Auger transition. When we apply the final state rule to the 
transition metal like Cu and Zn with a full filled d-band, the final 
states should be a localized atomiclike impurity states rather than 
the ground state of the metals. This kind of final states satisfy 
the requirement of Cini-Sawatzky theory for the occurrence of 
the atomiclike CVV Auger spectra naturally, because of the 
very small band width of the localized states. 
For the simplicity of the calculations, 
we use the initial states with a core-hole as the final states. 
The dynamical screening will cause changes in the details of 
the Auger spectral profiles we presented in Fig.10 and 11, but the 
left improvement between theory and experiment will not be 
dramatic or qualitative.

In conclusion, it may be stated that the occurrence of bandlike and
quasi-atomiclike features of the CVV Auger spectral line shape 
across the 3d and 4d transition metals can be correlated 
with the distinctly different behavior
of d-transition metals in screening a core-hole, the mechanism crucially
depending on whether their d-bands are partially or completely filled.

\acknowledgments
The author is very grateful to the very helpful discussion with 
L. Fritsche. This work was
supported by the National Science Fund for Distinguished Young Scholars 
under Grant No. 10025416 and also by the National Natural Science 
Foundation of China under 
Grant No. 19974075 and 59971064.\newline

\vskip 1.5 cm \newpage \noindent {\Large {\bf Figure Captions:}}\newline
\newline

\noindent{\bf FIG. 1.} Local density of states (LDOS) of Sc metal. 
The solid line refer to the atom in the center of the unit cell with a 
core-hole, the dot line to the atoms at the corner of the unit cell, 
and the long-short dashed line to the atom in a perfect crystal. The 
Fermi energy is chosen to be zero and indicated by the vertical thin 
long dashed line.

\noindent{\bf FIG. 2.} Same as Fig. 1 but for Ti.

\noindent{\bf FIG. 3.} Same as Fig. 1 but for Cr. Please note that a 
spin-dependent calculation has been performed to consider the 
antiferromagnetic property of the metal below $T_N$.

\noindent{\bf FIG. 4.} Same as Fig. 1 but for Fe. Please note that a 
spin-dependent calculation has been performed to consider the 
ferromagnetic property of the metal.

\noindent{\bf FIG. 5.} Same as Fig. 1 but for Ni. Please note that a 
spin-dependent calculation has been performed to consider the 
ferromagnetic property of the metal.

\noindent{\bf FIG. 6.} Same as Fig. 1 but for Cu. Please note that 
an additional $\delta$-function type structure is generated far below 
the original d-band for the atom with a $L_3$ core-hole, which 
indicates that localized physical impurity states are generated by 
the strong attractive core potential.

\noindent{\bf FIG. 7.} Same as Fig. 1 but for Zn. Please note that 
similar structure has been generated as in the case of Cu.

\noindent{\bf FIG. 8.} Same as Fig. 1 but for Pd. Please note that 
relatively sharp structure appears at the bottom of the d-band for 
the atom with a $L_3$ core-hole, but the band feature does not change 
anymore, reflecting the fact that Pd is not a full-filled d-band metal.

\noindent{\bf FIG. 9.} Same as Fig. 1 but for Ag. The existence of 
a core-hole affects the valence electrons in a similar way as for 
Cu and Zn.

\noindent{\bf FIG. 10.} The $L_3VV$ Auger spectra of Cu. The solid line 
refer to the results with the influence of the core-hole state, the 
dashed line to the results of a perfect crystal calculation.

\noindent{\bf FIG. 11.} The same as in Fig.10 but for the 
$M_1VV$ Auger spectra of Cu.

\newpage

\begin{center}Fig. 1\end{center}
\begin{figure}[htbp]   
\begin{center}
\setlength{\unitlength}{1truecm} 
\begin{picture}(6.8,6.8)
\put(-6.5,-20)
{\includegraphics{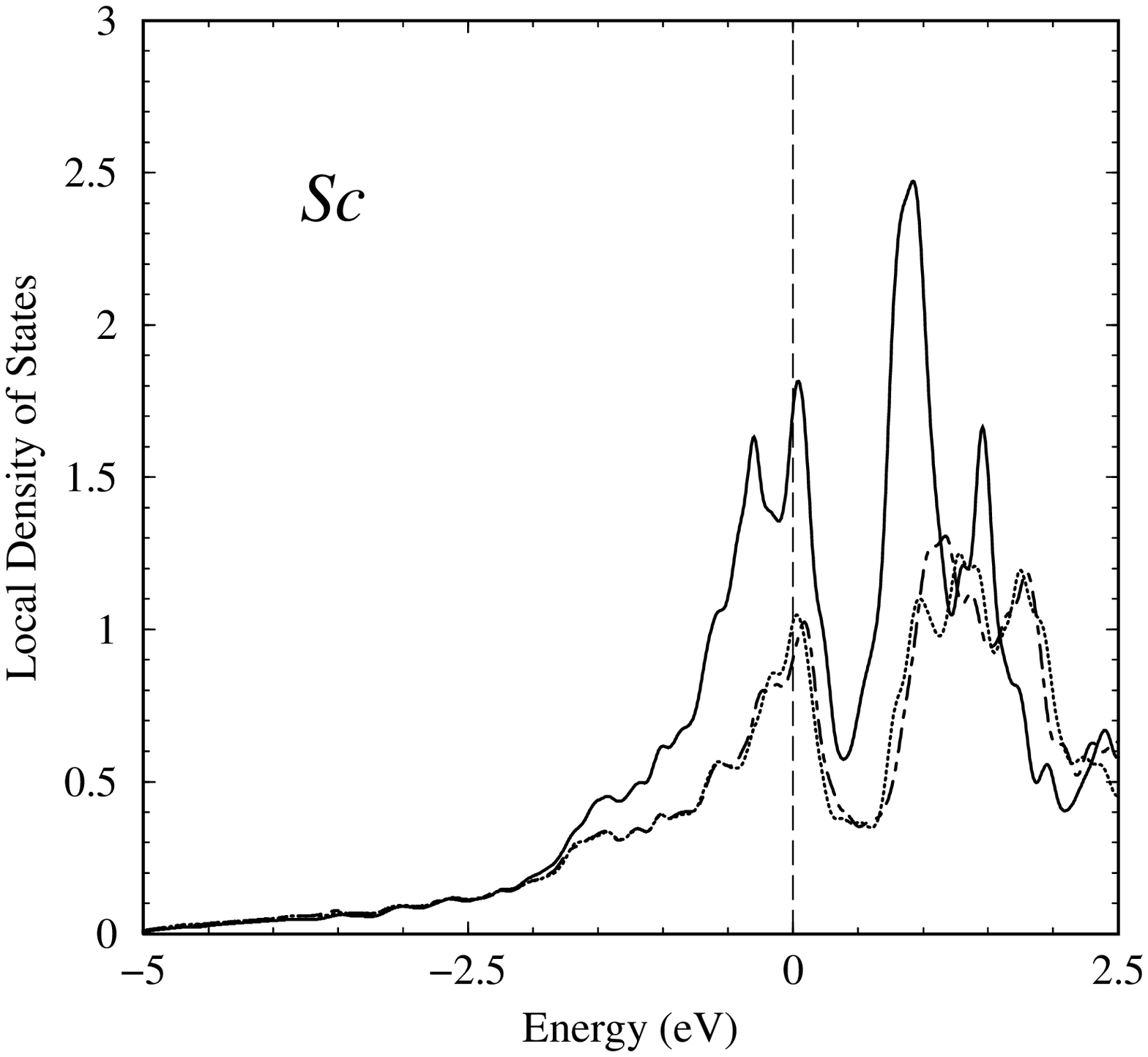}} 
\end{picture} 
\end{center} 
\vskip 2.0cm 
\protect\label{Fig.1}
\end{figure}
\hspace{5cm} 
\newpage
\begin{center}Fig. 2\end{center}
\begin{figure}[htbp]   
\begin{center}
\setlength{\unitlength}{1truecm} 
\begin{picture}(6.8,6.8)
\put(-6.5,-20)
{\includegraphics{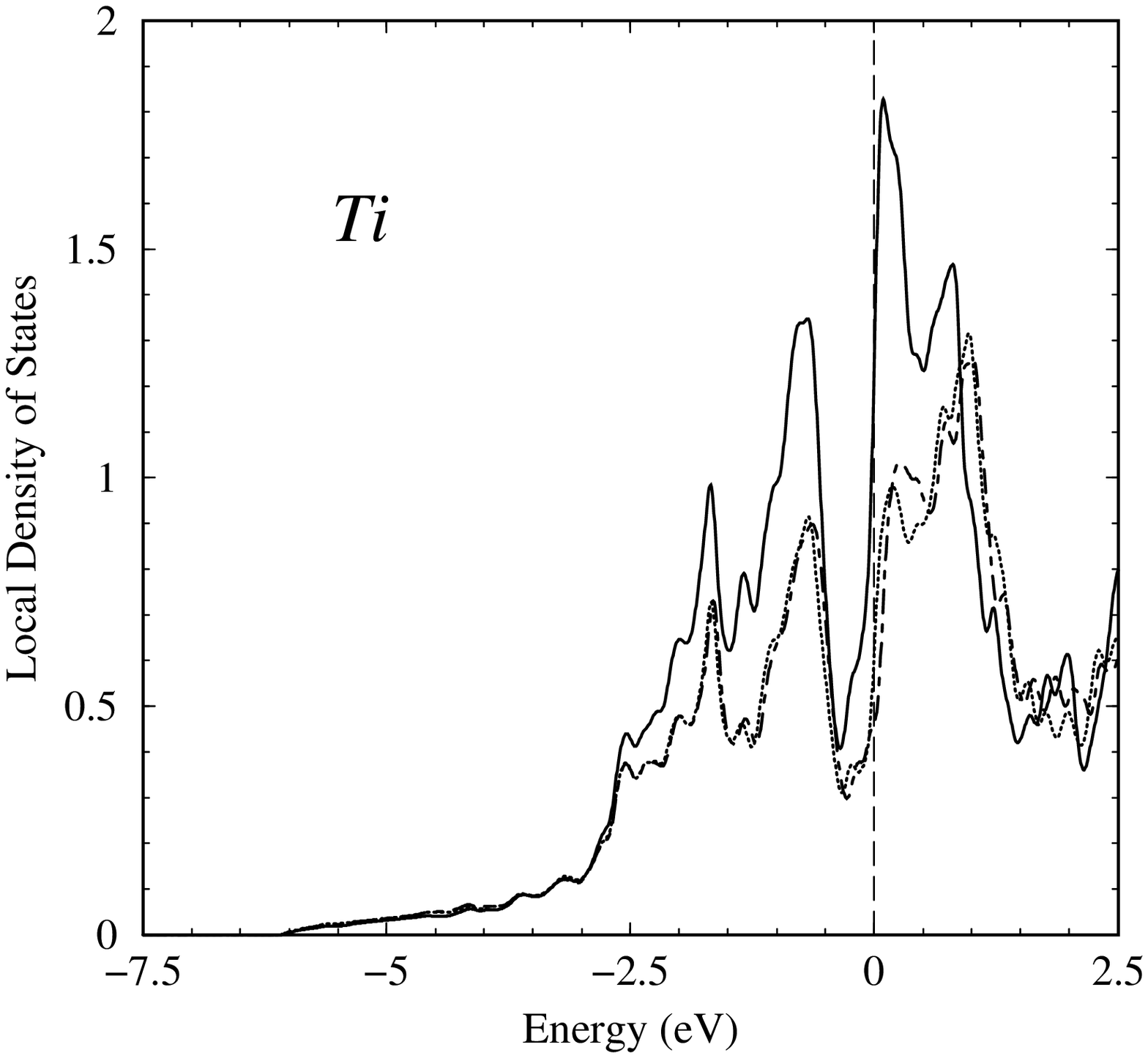}} 
\end{picture} 
\end{center} 
\vskip 2.0cm 
\protect\label{Fig.2}
\end{figure}
\hspace{5cm}
\newpage
\begin{center}Fig. 3\end{center}
\begin{figure}[htbp]   
\begin{center}
\setlength{\unitlength}{1truecm} 
\begin{picture}(6.8,6.8)
\put(-6.5,-20)
{\includegraphics{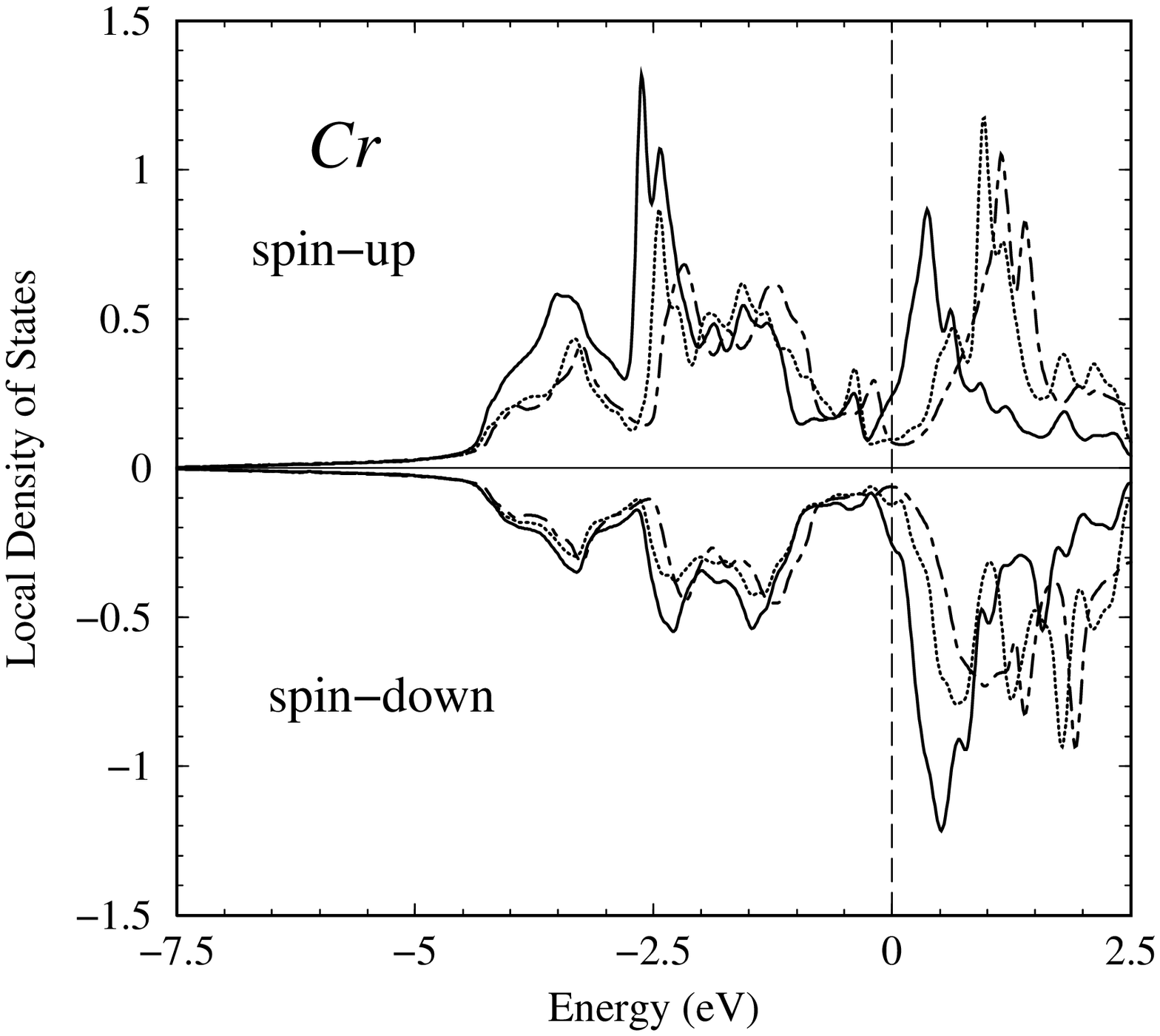}} 
\end{picture} 
\end{center} 
\vskip 2.0cm 
\protect\label{Fig.3}
\end{figure}
\hspace{5cm}
\newpage
\begin{center}Fig. 4\end{center}
\begin{figure}[htbp]   
\begin{center}
\setlength{\unitlength}{1truecm} 
\begin{picture}(6.8,6.8)
\put(-6.5,-20)
{\includegraphics{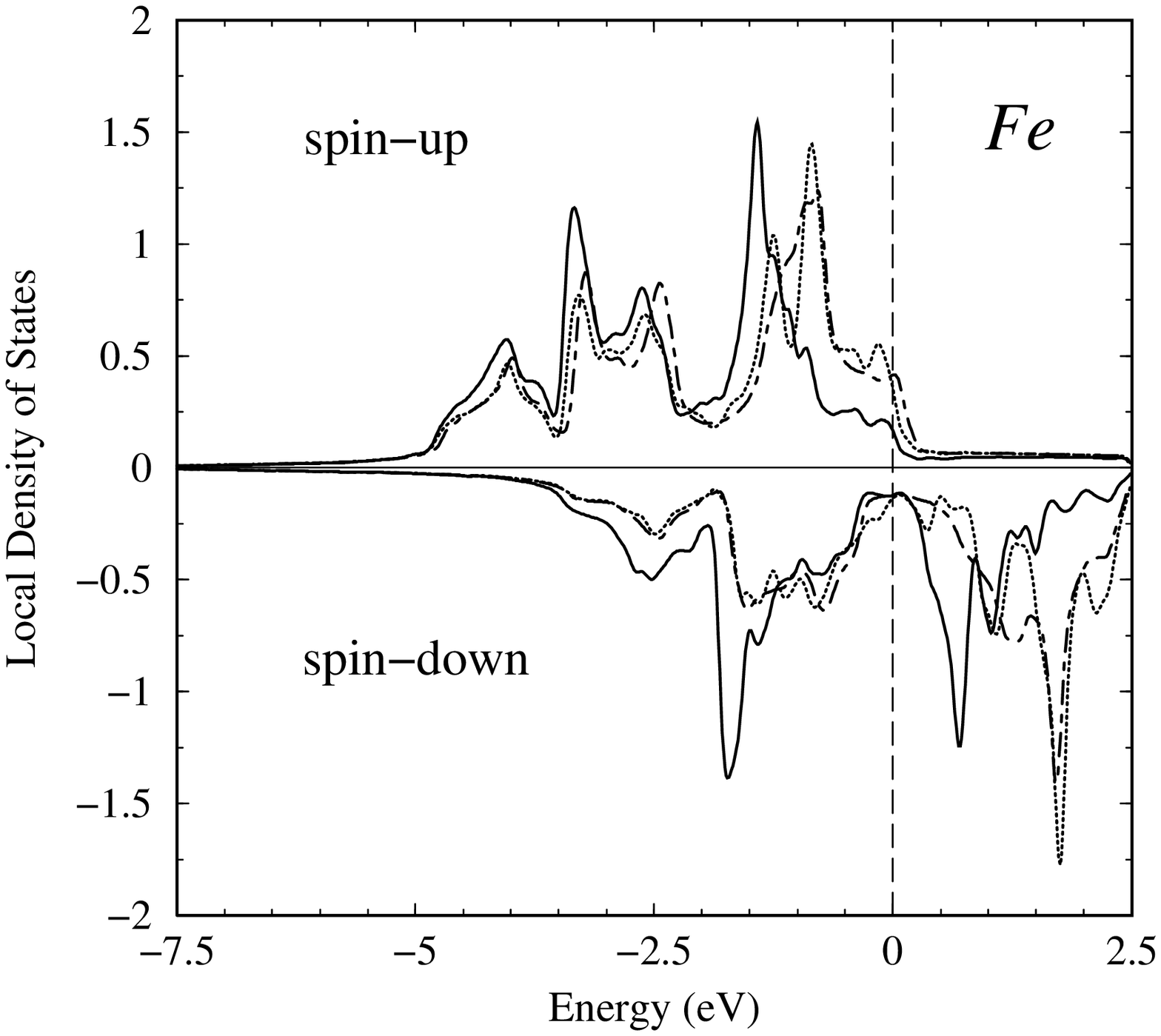}} 
\end{picture} 
\end{center} 
\vskip 2.0cm 
\protect\label{Fig.4}
\end{figure}
\hspace{5cm}
\newpage
\begin{center}Fig. 5\end{center}
\begin{figure}[htbp]   
\begin{center}
\setlength{\unitlength}{1truecm} 
\begin{picture}(6.8,6.8)
\put(-6.5,-20)
{\includegraphics{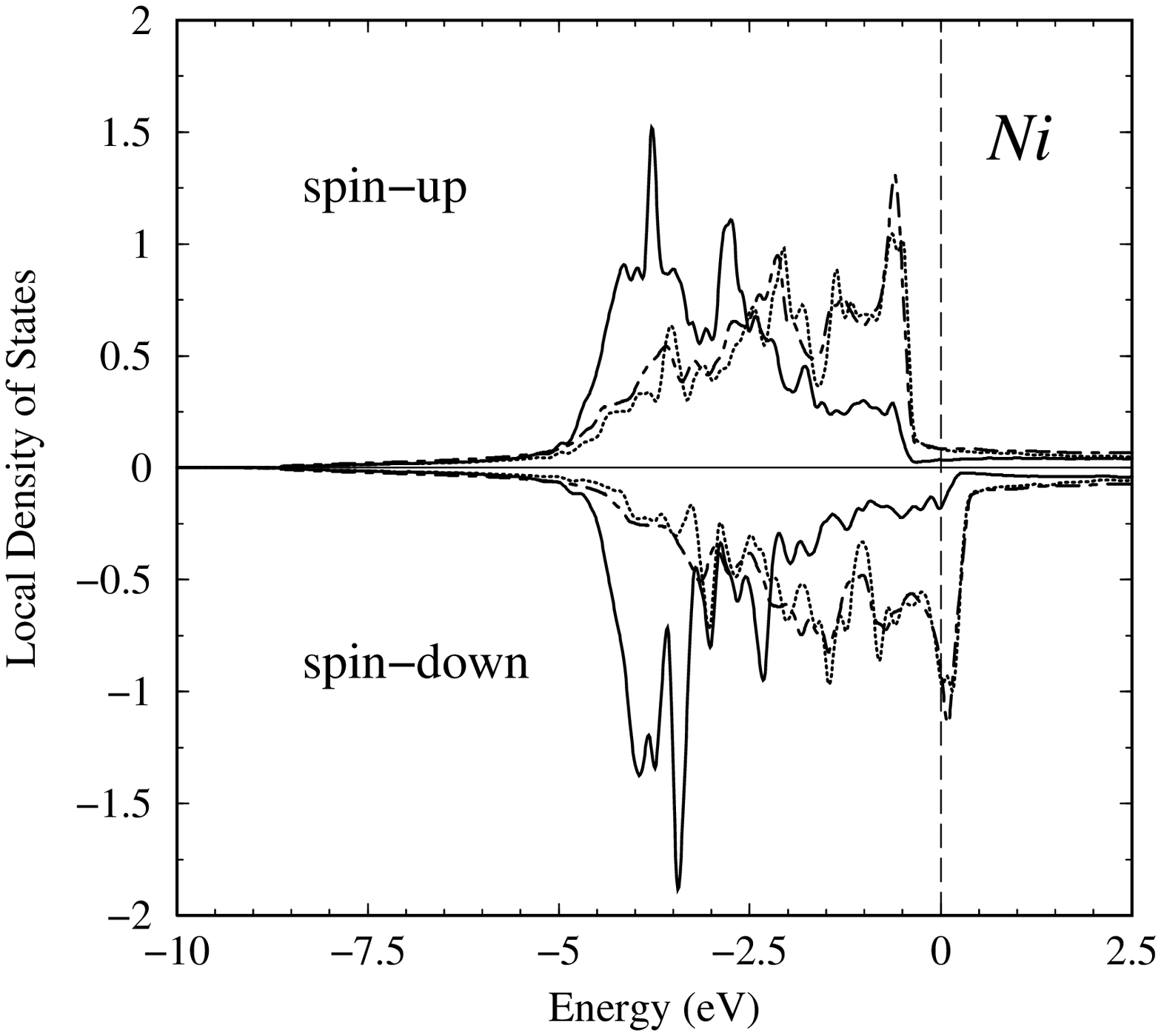}} 
\end{picture} 
\end{center} 
\vskip 2.0cm 
\protect\label{Fig.5}
\end{figure}
\hspace{5cm}
\newpage
\begin{center}Fig. 6\end{center}
\begin{figure}[htbp]   
\begin{center}
\setlength{\unitlength}{1truecm} 
\begin{picture}(6.8,6.8)
\put(-6.5,-20)
{\includegraphics{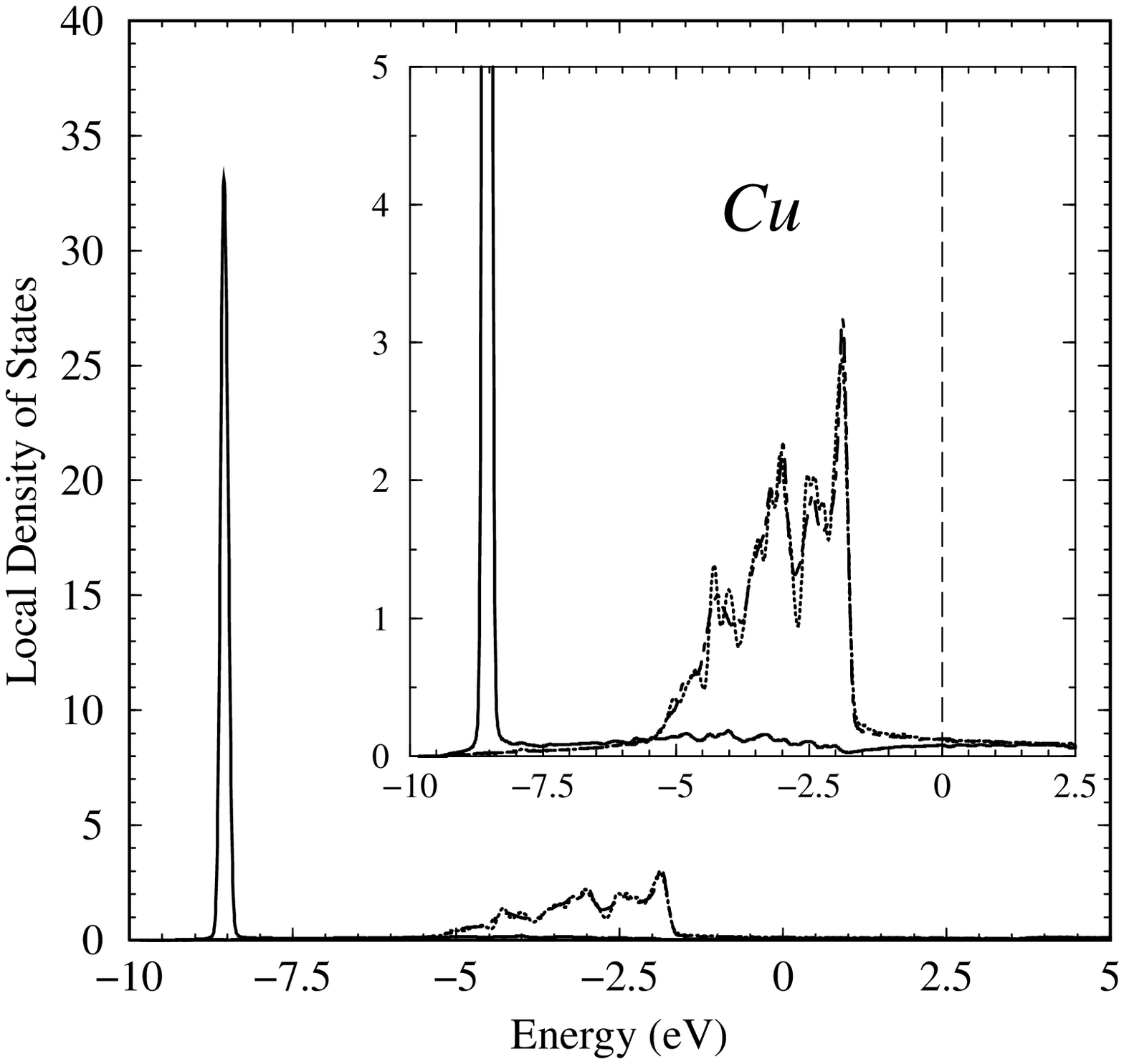}} 
\end{picture} 
\end{center} 
\vskip 2.0cm 
\protect\label{Fig.6}
\end{figure}
\hspace{5cm}
\newpage
\begin{center}Fig. 7\end{center}
\begin{figure}[htbp]   
\begin{center}
\setlength{\unitlength}{1truecm} 
\begin{picture}(6.8,6.8)
\put(-6.5,-20)
{\includegraphics{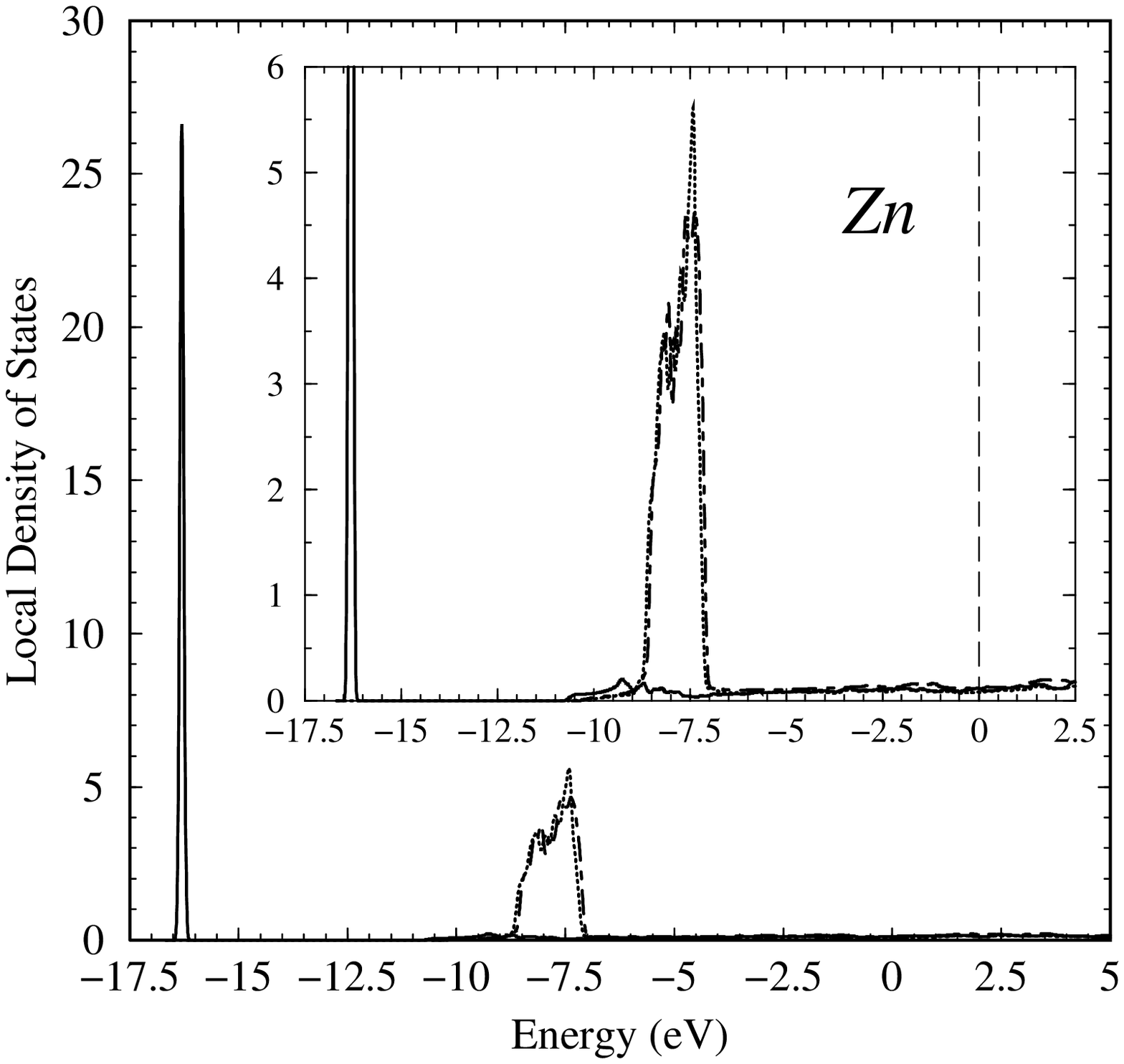}} 
\end{picture} 
\end{center} 
\vskip 2.0cm 
\protect\label{Fig.7}
\end{figure}
\hspace{5cm}
\newpage
\begin{center}Fig. 8\end{center}
\begin{figure}[htbp]   
\begin{center}
\setlength{\unitlength}{1truecm} 
\begin{picture}(6.8,6.8)
\put(-6.5,-20)
{\includegraphics{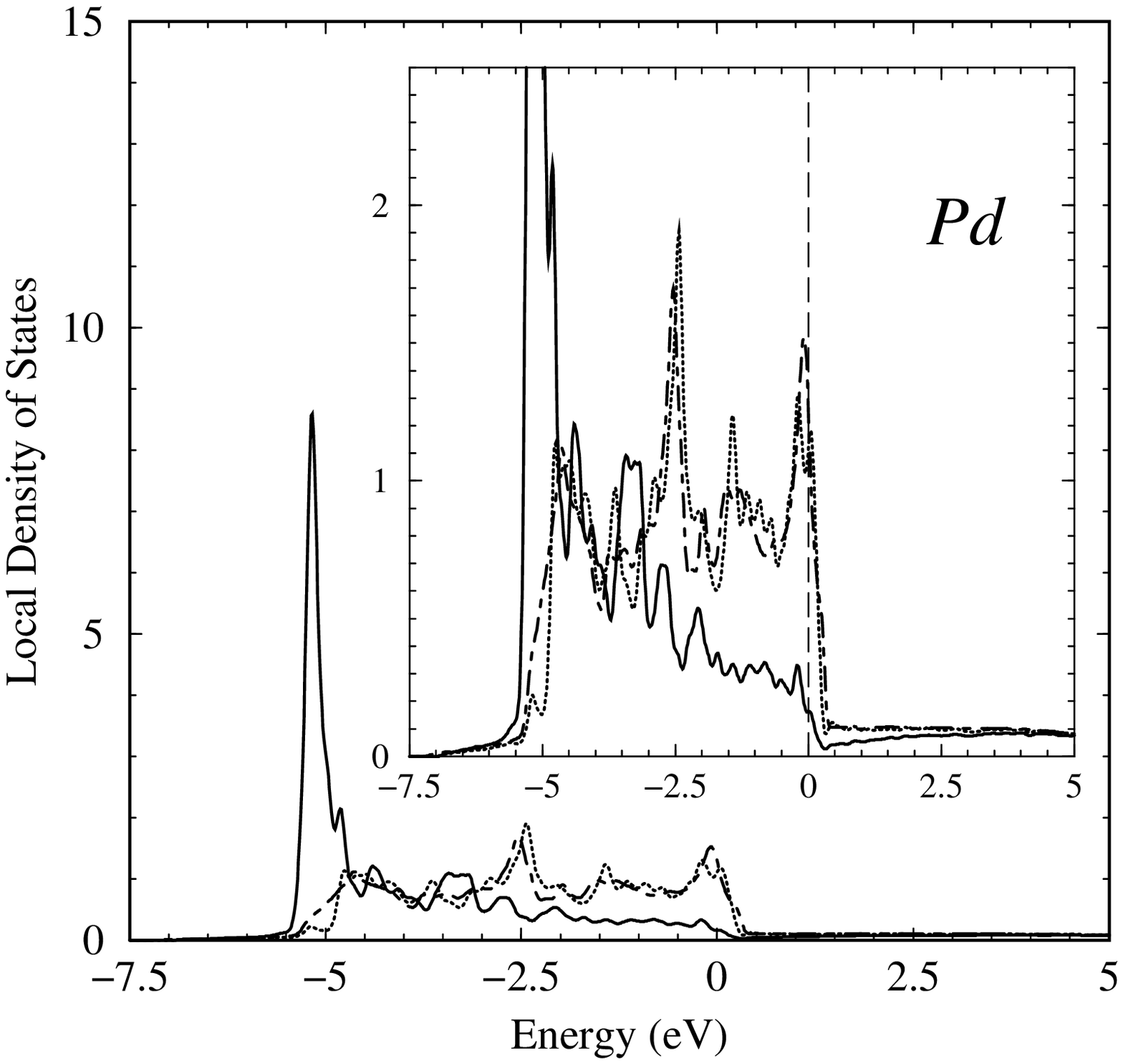}} 
\end{picture} 
\end{center} 
\vskip 2.0cm 
\protect\label{Fig.8}
\end{figure}
\hspace{5cm}
\newpage
\begin{center}Fig. 9\end{center}
\begin{figure}[htbp]   
\begin{center}
\setlength{\unitlength}{1truecm} 
\begin{picture}(6.8,6.8)
\put(-6.5,-20)
{\includegraphics{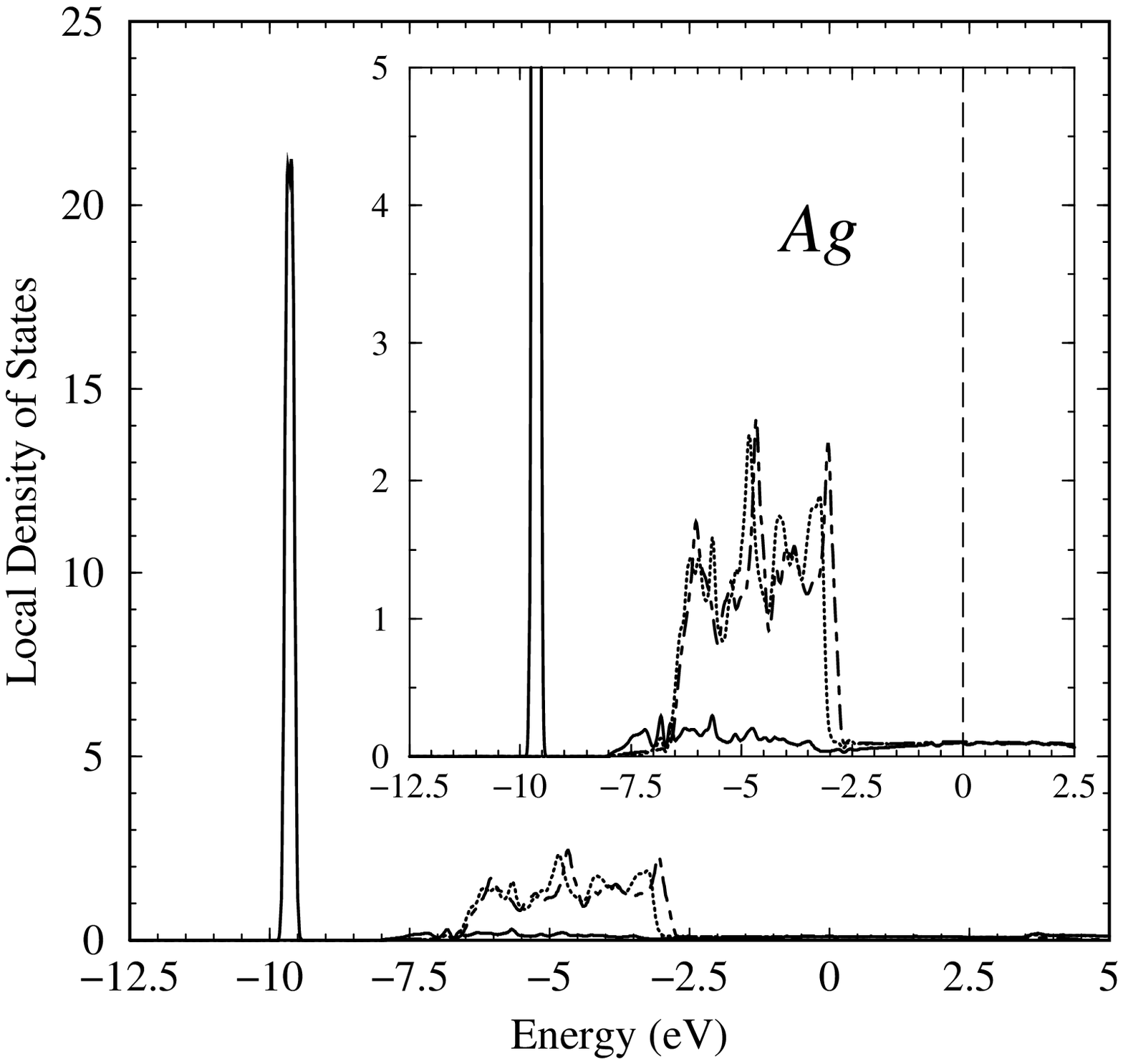}} 
\end{picture} 
\end{center} 
\vskip 2.0cm 
\protect\label{Fig.9}
\end{figure}
\hspace{5cm}
\newpage
\begin{center}Fig. 10\end{center}
\begin{figure}[htbp]   
\begin{center}
\setlength{\unitlength}{1truecm} 
\begin{picture}(6.8,6.8)
\put(-6.5,-20)
{\includegraphics{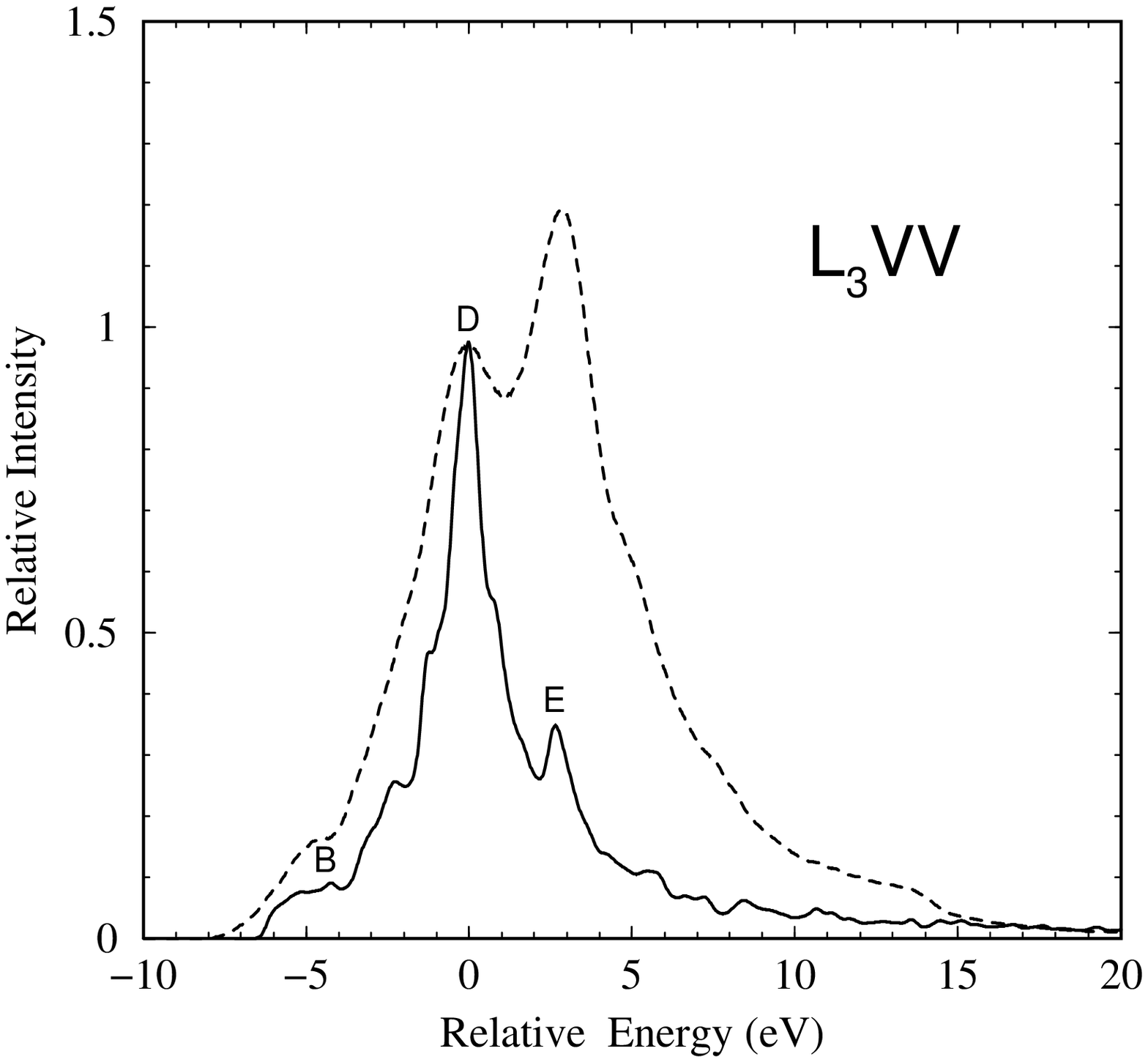}} 
\end{picture} 
\end{center} 
\vskip 2.0cm 
\protect\label{Fig.10}
\end{figure}
\hspace{5cm}
\newpage
\begin{center}Fig. 11\end{center}
\begin{figure}[htbp]   
\begin{center}
\setlength{\unitlength}{1truecm} 
\begin{picture}(6.8,6.8)
\put(-6.5,-20)
{\includegraphics{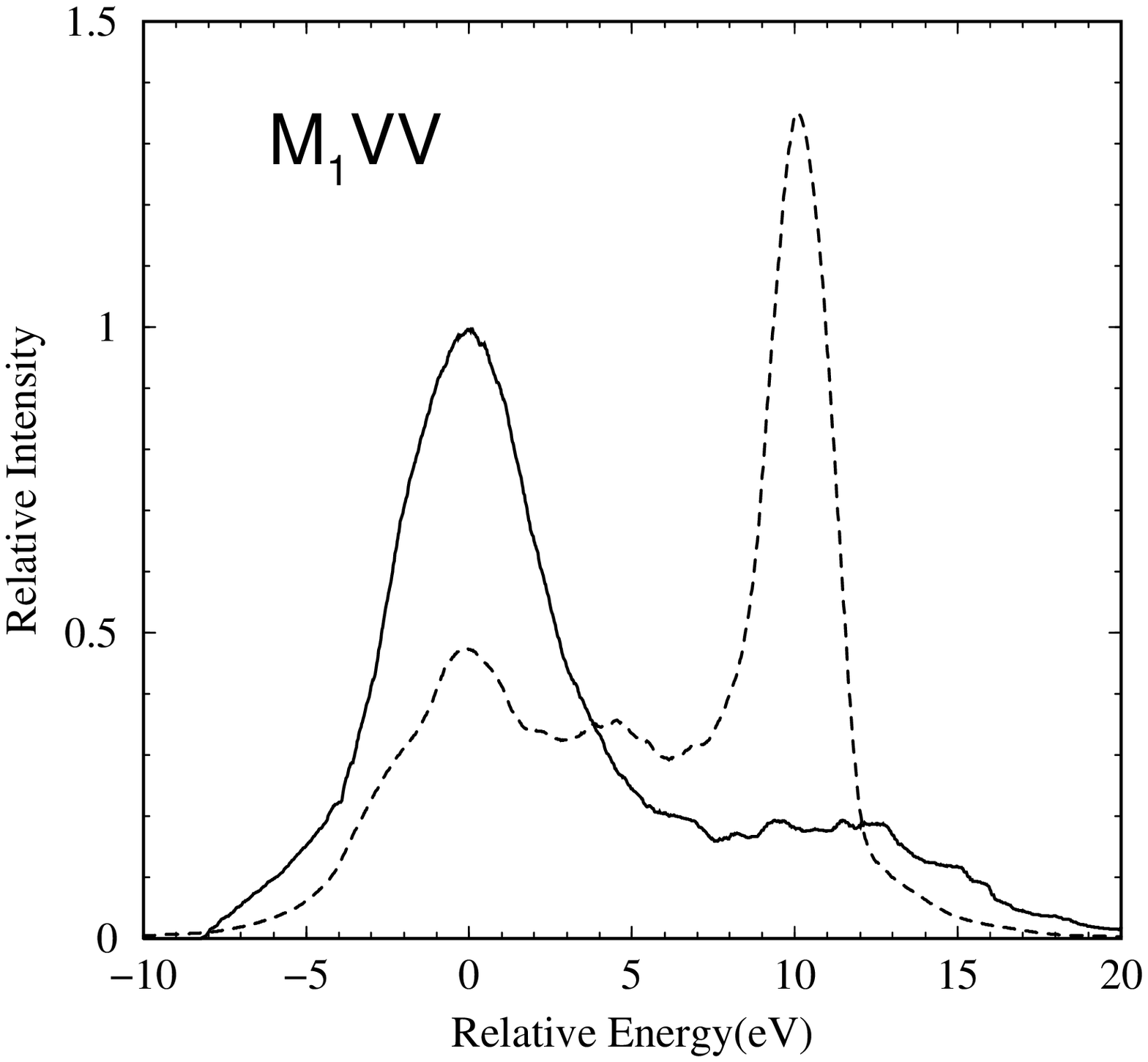}} 
\end{picture} 
\end{center} 
\vskip 2.0cm 
\protect\label{Fig.11}
\end{figure}
\hspace{5cm}

\end{document}